\documentclass[11pt, a4paper]{article}
\begin{document}
\title
{Anomalous temperature dependence of surface tension and capillary waves at 
liquid gallium surface }
\author
{V. Kolevzon\\
Institute for Physical Chemistry, University of Karlsruhe,\\
76128 Karlsruhe, Germany}
\maketitle
The temperature dependence of the surface tension $\gamma(T)$ at liquid
gallium is studied
theoretically and experimentally using light scattering from capillary waves.
The theoretical model based on the 
Gibbs thermodynamics relates  
$\partial\gamma/\partial${\it T} to the surface excess entropy density $-\Delta S$.
Although capillary waves contribute to 
the surface entropy with a positive sign, the effect of dipole layer
on $\Delta S$ is negative. 
 Experimental data collected at a free Ga surface 
in the temperature range from 30$^\circ$ C to 160$^\circ$ C show 
that $\partial\gamma/\partial T$ changes sign near 100 $^\circ$C.
\section{Introduction}
\noindent
The temperature dependence of the surface tension of liquids  is 
very important in applications such as Marangoni convection and
crystal growth. However, little is known about the specific
surface forces influencing this dependence in liquid metals.
      
Our previous works \cite{kol, kol1} were concerned with the surface 
tension of mercury as a function of temperature, which was studied using
quasielastic light scattering from capillary waves \cite{lang}. 
Although the experimental data show a decrease in surface 
tension with {\it T}, the theoretical explanation of this fact seems to
be non trivial. 
The recent discovery of surface layering in liquid
gallium and mercury indicates an ordered metal structure 
perpendicular to the surface which is about a few atomic 
diameters thick \cite{reg,mag}.
This ordering can drastically influence
the entropy density profile in the surface zone, reducing the 
entropy density at the surface compared to the entropy 
density of the bulk. It was suggested \cite{kol}
that a liquid metal surface is a two-component system comprising quasi--
free surface electrons and positive ions. In this case
$\partial\gamma/\partial T$ 
would depend on the surface excess entropy as well as
the chemical potential of surface electrons \cite{kol}.
However, the ionic and 
electronic profiles of any complicated shape should be approximated by the 
same right-angled 
profile \cite{kol1} in order to preserve the electrical neutrality of metal as 
a whole.
Thus, the two-component model is inapplicable to a free metal surface,  
whose excess entropy should be evaluated in the framework of 
a one-component model \cite{land,crox}
\begin{equation}
\label{entrop}
\Delta S=-\frac{\partial \gamma}{\partial T},
\end{equation}
where $\Delta S$ is defined
$\Delta S=\int_{-\infty}^{\infty} dz(S(z)-S_b \theta(z))$.
Here $\theta$(z)=0 for z$>0$ and 
$\theta$(z)=1 for z$\leq 0$; z=0 denotes the surface position
and the integration is performed from the liquid bulk (z=-$\infty$) 
to a vapor phase (z=$\infty$) and S$_b$ denotes the entropy density in
the liquid bulk.
We split the surface excess entropy into three parts describing surface
layering, capillary waves, and a surface double layer \cite{kol1}. 
As expected
the contribution from surface layering  into $\Delta S$ is negative 
\cite{kol},
whilst that contribution from capillary waves is positive. It was suggested
\cite{kol1} that the electric field in the double layer would contribute 
to the surface excess
entropy with a positive sign if the layer thickness varies with T but
the surface charge density does not.
The present paper demonstrates that this is not the case; the thickness of
the dipole layer remains insensitive to the temperature variations  
 and the contribution
from electric field into $\Delta S$ should be negative indicating a decrease
in the surface electron density on heating. 
  
\section{Theoretical estimates of $\Delta S$}
                           
A liquid metal comprises two components: free electrons 
and positive ions. Free electrons behave as a quantum medium: even at 
zero temperature their energy is nonzero, and is usually written via 
the Fermi energy E$_f$: $E=E_f$N, where N is the number of electrons.
The spatial electronic density profile $\rho_e(z)$
has a nonzero width (even at 0K), that leads to some redistribution of 
electrons between
bulk and the surface. In other words, some electrons are ejected from the
bulk and concentrate on the vapor side of the  interface.
This charge separation leads to a surface double layer,
suggested by Frenkel \cite{fren}. 
                           
Unfortunately, no theory describing the surface tension
of liquid metals is well established. Numerical simulations of the 
electronic and ionic density profiles \cite{rice} 
predict oscillations of the density profile but the driving force
of these oscillations is not very clear. 
Only one paper, to the author's knowledge, treats this problem analytically
\cite{samo}. Due to its importance for the present context we 
repeat the main results of this paper. 

A detailed analysis done by Samojlovich at 0 K
 ensures that the electrostatic force $-\rho_e\nabla\phi$
leads to Maxwellian elastic stresses
$\Pi_{zz}=-\Pi_{xx}=\epsilon_0 E^2/2$ 
(z axis is directed along the surface normal and 
x--along the  surface)
or electrostatic pressure P=$\epsilon_0 E^2/2$ deforming the ionic fluid. 
This isotropic pressure acting on the topmost layer makes the atoms there  
over-compressed compared to atoms in the bulk.   
These considerations predict a non-monotonic density
profile with a maximum in the topmost  
layer, which was demonstrated recently on the 
surface of liquid gallium and mercury.
 
The surface tension of any liquid 
can be expressed via the tensor of anisotropic stresses with the
components $\Pi_{zz}$ and $\Pi_{xx}$  appearing as a result of the 
density gradient in the surface zone. The surface tension of a 
liquid metal at 0K is given by 
\begin{equation}
\label{tension}
\gamma=\int_{-\infty}^{\infty}dz (\Pi_{zz}-\Pi_{xx})=\int_{-\infty}
^{\infty}dz [\chi_w
\frac{1}{\rho}(\frac{\partial\rho}{\partial z})^2-\epsilon_0 E^2]
\end{equation}
where $\chi_w=\frac{\hbar^2}{4m}$ \cite{samo}.
This expression is obtained from the balance between the surface 
forces, namely the electrostatic force
 $-\rho\nabla\phi$ and the quantum pressure of electron gas given by
von Weizsacker's term $\chi_w\int dz\frac{1}{\rho}(\nabla\rho)^2$.
If only the electrostatic energy were accounted for the surface
tension would be negative \cite{samo}.
In order to calculate the surface tension one should know the electron
density profile and the solution of Poisson's equation in the surface
zone. 
 
The temperature behavior of the surface tension of a liquid metal 
is a most intriguing problem. While the changes in tension for nearly
all liquid metals, as expected
from literature, are only few percent over 100 K \cite{gal},
the temperature derivative
$\partial\gamma/\partial $T should be extremely sensitive to the 
density profile \cite{crox}. 
In the following text we will show some basic relations
connecting the surface excess entropy to other physical properties
of the surface layer. 

It is supposed that the surface entropy 
comprises different parts \cite{crox}: the first, pertinent to all fluids, 
is due to capillary waves. The second part is an orientational
entropy $\Delta S_{or}$ determined by
surface layering. As we discussed in \cite{kol} the variation of the entropy
should have a minimum in the surface zone because the surface atoms are
more ordered than atoms in the bulk. This would require 
$\Delta S_{or} < 0$ as it was schematically drawn in  \cite{kol}.
There might be an orientational contribution in $\Delta S$ due to 
ordering in the surface plane.
This ordering, which would appear in the form of a hexatic
monolayer on isotropic liquid substrate, is responsible for a change in
sign of the temperature derivative of the surface tension of normal alkanes
and some liquid crystals \cite{alkan,crox}.
For liquid metals in-plane surface order was
never reported in experiments; only atomic ordering
along the surface normal was detected \cite{reg,mag}. Also from a theory
this hexatic layer on a pure liquid metal cannot be expected: for the
formation of the Wigner lattice one needs a very low electron density \cite
{wign}.
We show below that the impact of surface layering on $\Delta S$
is naturally linked to the specific surface effects namely the anisotropic
pressure and the dipole layer.
 
According to Frenkel's analysis the surface excess entropy is \cite{fren}
\begin{equation}
\Delta S=n_s k_B \ln\frac{\bar{\omega}_p}{\bar{\omega}_c}.
\end{equation} 
where n$_s$ is the number of atoms per unit area, 
$\bar{\omega}$ is the mean frequency of capillary 
waves, defined as
$\bar{\omega}_c=\frac{1}{n_s}\int_0^{q_{max}}dq\omega(q)2\pi q$ and
$\bar{\omega}_p$ is the mean frequency of bulk phonons.
$\Delta S$ is governed by the ratio of two mean 
frequencies surface and bulk and is positive because $\bar{\omega}_p >
\bar{\omega}_c$.  
In order to calculate $\bar{\omega}_c$ one should know the dispersion
relation $\omega(q)$ in the whole q-range: from long waves to the shortest
wave limit. It should be borne in mind that the well-known relation
$\omega\sim q^{3/2}$ derived from a linear theory of capillary waves
 is valid only for long waves, ie. relatively
small q. For high wavenumbers, giving the main contribution in
$\bar{\omega}_s$, this relation is not valid any more because
linear theory fails at some q such that the wavelength of the wave
is of the same order as the amplitude. Hence it is not possible to
 calculate correctly the contribution of capillary waves in $\Delta S$
at least in the framework of a phenomenological approach.
                                                            
Now we find out the impact of surface
forces on the surface entropy which is specific to all metalls.
These are the electrostatic force and anisotropic pressure of electron
gas in the surface zone.
The contribution of electric field to the surface free energy is 
$F_s \propto -\epsilon_0 E^2\delta z$, as discussed above.
We suggest that the structure of expression for the free 
energy Eq(\ref{tension}) of a metal
surface given at 0K is valid for T$\neq 0$ but
the two terms in Eq(\ref{tension}) are supposed to be temperature dependent.
Then the surface excess entropy should be affected by 
the electrostatic energy and the gradient of the electron density;
both are specific solely to the surface zone and are
zero in the bulk metal. The standard expression $S=-\partial F/
\partial T$ yields for the entropy
\begin{equation}
\label{es}
\begin{array}{r}
\Delta S_{es}=-\frac{\partial}{\partial T}\int_{-\infty}^ {\infty}dz
(\chi_W[\frac{1}{\rho}(\frac{\partial\rho}{\partial z})^2]
-\epsilon_0 E^2)=\\ \\
=-\delta z\frac{\partial}{\partial T}
[\frac{\chi_W}{\rho}(\frac{\partial\rho}{\partial z})^2]
+2\epsilon_0\mid E\mid \delta z\frac{\partial E}{\partial T}  
\end{array}       
\end{equation}
Here we make an essential assumption
that the characteristic length scale $\delta z$
of the surface zone remains unchanged with T whereas the
squared density gradient and electric
field should vary. This is so because the spatial density of surface electrons
varies rapidly on the scale $\sim k_F^{-1}$ as the Fermi wavelength k$_F$
is T-independent. By contrast, the amplitude of the first peak 
in the electronic density $\rho(z)$ is temperature sensitive, 
at least over T-range from 30 to 200 $^\circ$C \cite{reg1}. 
According to Samojlovich's analysis \cite{samo} we associate
the appearance of this maximum with the strength of surface electric field
that should decrease on heating as the layering amplitude does \cite{reg1,rice}.
Put in other way,
the density of over-compressed surface atoms decreases due to a decrease
in the density gradient and electrostatic pressure, whilst the thickness of
double layer is nearly constant.
The sign of $\Delta S_{es}$ depends on the interplay between the two  
terms in Eq(\theequation) 
whose numerical amplitudes remain essentially unknown to the present 
study. Insofar as $\partial E/\partial T$ is negative  $\Delta S_{es}$,
 may be negative too. Due to the opposite effects produced by the density gradient
and the electric field the amplitude of $\Delta S_{es}$ is expected 
to be small, hence $\partial\gamma/\partial T$ is mainly affected
by capillary waves especially at elevated T.
However the theory outlined points out
a possible negative sign in the surface excess entropy in liquid metals  
when  $\Delta S_{es}$ overcomes capillary wave entropy $\Delta S_{cap}$.                                   
\section{Experimental methods}              
Our light scattering technique is described in detail
elsewhere \cite{lang,earn5,kol2}. In brief, a beam from a 7 mW He-Ne laser
(TEM$_{00}$, $\lambda$=632 nm) fell on the liquid surface. 
Small-angle scattered light was optically mixed (on a 
photodetector) with a portion of the original beam, providing all the
necessary conditions for optical heterodyning. The output of a photomultiplier 
was modulated at the propagation frequency of a capillary wave with 
the selected wavenumber q. The spectral representation of the signal was 
recorded in
the frequency domain with a spectrum analyzer. The whole
apparatus was placed on an optical table, vibration
isolation being provided by four pressurized air cylinders  
in the legs. 

The liquid Ga surface was prepared in a high vacuum chamber 
 (10$^{-7}$ to 10$^{-8}$ Torr) by dropping
 Ga from a glass cylinder terminating in a capillary of diameter less
 than 1 mm. Liquid Ga (purity 99.999, Alfa) was dropped in vacuum to a
 Ti trough having a
diameter 50 mm. The low melting rate ensures that a thick
 film of oxide remains on glass walls while pure metal flows through
 the capillary. Usually it takes some hours to complete the layer with
 strong curvature near the walls and a flat area in the middle part. 
Though no visible
 traces of contamination on the liquid surface were detected by eye the
 laser beam displayed a halo caused by scattering from a transparent 
oxide film which had flowed through the capillary despite all precautions. 
 This continuous oxide film was fired
 with an Ar-ion sputter gun having the beam energy of 2keV
 at 10$^{-5}$ Torr pressure of Argon. Although the ion beam hits
 the area smaller than the whole surface,  the film particles flow
 along the gradient of surface tension to the beam footprint.
 Hence a self cleaning mechanism due to Marangoni flow works on the liquid surface
 sputtered in vacuum. The footprint of a laser beam on the sputtered
 gallium surface was hardly visible indicating the mirror
 finish and the absence of particles on the liquid surface (apart from an
island of oxide particles sticking to the wall). Note that the experimental
$\omega$(T) dependencies reported below were obtained with the sputter gun on.
Switching off the ion source caused an increase in the visibity of the laser
beam footprint due to enhanced scattering from an oxide layer formed either
due to oxidation by the residual gas or due to re-distribution of
oxide particles on the Ga surface.

 Due to very high surface tension and non-wetting between gallium and
 the container a Ga-layer, usually 6-8 mm thick in
 the middle, easily supports the propagation of vibrations in the liquid sample.
 Although the whole chamber was placed
 on an optical table, mechanical disturbances from a turbo-molecular pump 
 caused such high vibrations of the liquid sample that no wave spectra could
 be measured. We tried to obtain wetting between liquid Ga and the container
 in order to reduce the liquid depth. Wetting can be easily
 obtained by heating the sample in vacuum above 500 C. At this temperature
 Ga spreads over the container surface and thin layers can be prepared.
 However the container material very likely dissolves 
in gallium at high temperature as follows from our experiments. 
To keep the liquid 
sample clean one should try to achieve wetting at room temperature.

 In general, liquid gallium dropped in vacuum wets a polished container
 wall in some places but it spreads neither along the wall
 nor along the bottom. Therefore a Ga drop should be rolled along
 the walls until a ring of liquid is formed.  Then the layer could be
 completed by shaking the liquid sample. After the layer is completed
 its form is determined by the attraction potential between Ga and walls;
 usually the surface profile was a highly non-symmetric (along azimuth)
 with concave and convex parts. Only partial wetting could be 
achieved: places that were wetted had a strong meniscus along the wall upwards;
 where no wetting took place the liquid edge was directed downwards. 
The Ga-layer  prepared in such a way  is about 3 to 
4 mm thick in the middle. 
Due to a reduction in amplitude of vibrations the wave spectra could 
be collected on such samples; wetting helps to achive shallow and 
tense layers.    

Capillary waves scatter light mainly at small angles about the 
reflected beam. The spectrum of the scattered light is the 
power spectrum of capillary waves, which is approximately
Lorentzian \cite{lang}. The data were fitted with
a theoretical function that incorporates
the effects of instrumental broadening \cite{hard,earn0}. 
The spread $\delta q$ in the wavenumbers gives a 
corresponding broadening $\Delta\omega$ in the spectrum. For the  
Gaussian beam the instrumental function is also a Gaussian 
\cite{hard,earn0}. The convolution of an ideal Lorentzian and the Gaussian 
instrumental function of width $\beta$ yields \cite{earn0} 
\begin{equation}
P(\omega)=\int_{-\infty}^{\infty} \frac{\Gamma/\beta
\exp[-(\omega-\omega')^2/\beta^2]}{\Gamma^2+(\omega'-\omega_0)
^2} d\omega'.
\end{equation}
This integral can be evaluated in terms of the complementary
error function of the complex argument \cite{hard,abra}:
\begin{equation}
\begin{array}{r}
\label{fit}
S(\omega)=A\Re[\exp (-(i\Gamma/\beta+(\omega-\omega_0)
/\beta)^2)\\ \\ 
 {\rm erfc} (-i[i\Gamma/\beta+(\omega-\omega_0)/\beta])]+B,
\end{array}
\end{equation}
where A is the scaling amplitude and B the background. 
Thus, five properties were extracted from the fit of 
experimental spectra: frequency $\omega_0$, damping 
constant $\Gamma$, instrumental width
$\beta$, amplitude A, and background B. In the present context we concentrate
only on peak frequencies $\omega_0$.  
\section{Results}
The temperature behavior of peak frequencies $\omega_0$ 
of capillary waves at the free gallium surface is shown in Fig. 1.
To first order, the roots of the dispersion
relation describing the propagation of a capillary wave with
a particular wavenumber q are \cite{lang}
\begin{eqnarray}
\label{root}
\omega_0 & = & \sqrt{\gamma q^3/\rho}, \\
\Gamma     & = & 2\eta q^2/\rho, 
\end{eqnarray}
where $\gamma$, $\eta$, and $\rho$ are the surface tension,
bulk viscosity and density, respectively.
Equation (\ref{root}) serves as a basis for evaluation of the
tension. The data in Fig. 1 are fit by assuming 
a linear temperature dependence of the surface tension,
$\gamma=\alpha(T-T_0)+\gamma_0$, where $T_0$ is the melting point:
\begin{equation}
\label{slope}
\omega_0(T)=\sqrt{(\alpha(T-T_0)+\gamma_0) q^3/\rho}.
\end{equation}
The best-fit estimate of the slope of the temperature dependence
in the temperature interval from 100 $^\circ$C to 160 $^\circ$C is:
$\alpha=d\gamma/dT=-0.57 \pm$0.09 (mN/m)/K. 
The best estimate of $\gamma_0$
corresponding to the tension at the melting point 
(30$^\circ$ C) is $\gamma_0$=766 $\pm$17 mN/m.

In the temperature range from 30 to 100 $^\circ$C the temperature
dependence is anomalous and the best estimate of the slope
is $\alpha=4.18 \pm 0.2$ (mN/m)/K and the tension at the melting point
is $\gamma_0$=690$\pm$6 mN/m. 
The two data sets at temperatures above 100 C and below it were collected 
on the same sample but at two different q (295 and 310 cm$^{-1}$) which was
dictated by the necessity of optical adjustment of the reference beam.
Theoretically there is no limitation for a 
negative slope (at T far away from the critical point) but there is one for 
a positive $\alpha$. The point is that a negative
excess entropy indicates that the entropy density at the surface is less
than that density in the bulk. We do not know how much
this density changes; however we can estimate a maximum variation of
the entropy whose spatial profile cannot fall below zero in the surface 
zone (for graphic interpretation see \cite{kol}).
For liquid Ga at 303 K we find the bulk entropy per mole 
S= 9.82+4.4=14.2 cal/mol/K \cite{hand}.
Then the entropy density is $S_b=S \rho/M_{Ga}$=
59.7$\times$6.1/69.7=5 MJ/m$^3$/K, where $M_{Ga}$ is the molar weight
of Ga. In order to obtain the
surface entropy density this value should by multiplied by the thickness
of surface zone: $(S_s)_{max}=S_b\delta z=5\times 10^{6}\times 10
\times 10^{-10}$=5 mJ/m$^2$/K
where $\delta z \sim$10 {\AA} was suggested to be the characteristic
length scale of oscillations of the atomic density \cite{reg}.   
This value confirms
that the variations of the entropy occur in a relatively thick surface
zone of about 10 {\AA} indicating that the electrostatic interactions
penetrate deeper than the topmost atomic layer. $\alpha$=
4.18 mN/m/K extracted from experimental data at q=295 cm$^{-1}$ is in 
accord with 
$\alpha$ =3.87 mN/m/K obtained from peak frequencies at 456 cm$^{-1}$
on the surface of another sample
(Fig. 1 (b)) demonstrating an increase in 
surface tension on heating as a real fact and not due to vibrational
changes in $\omega$. The damping constants inferred from the fit were
strongly influenced by vibrations: the disperisons of $\Gamma$ were
up to 100 percent of the mean values. Therefore no quantitative 
predictions of the magnitude and
the temperature behavior of $\Gamma$ could be made in the present experiment.

The $\gamma(T)$ behavior can discussed now. 
The effect of capillary waves on $\Delta S$ is positive and $\Delta S_{cap}$
is very likely T independent. On the other hand $\Delta S_{es}$ becomes
negative according to Eq(\ref{es}) when the condition $\epsilon_0 E
\mid\frac{\partial E}{\partial T}\mid >\chi_W\frac{\partial}{\partial T}
[\frac{1}{\rho}(\frac{\partial\rho}{\partial z})^2]$  is fullfiled.
This means that both the density gradient and the electric field decrease
on heating (because a part of the surface electrons is injected into the bulk)
setting up $\partial\gamma/\partial T<0$ untill
$\mid\Delta S_{el}\mid>\Delta S_{cap}$ is valid.
Of course, this mechanism drives the magnitude            
of $\Delta S$ untill the Fermi distribution of electrons
 is T independent; for metals it is valid up to 1000 K.   
                            
The light scattering experiment does not directly yield the 
surface tension; only the peak frequency is directly measurable. 
In order to obtain additional information on the sign of
$\partial\gamma/\partial T$, the following experiment was
carried out. A temperature gradient imposed along the
sample  causes a liquid metal flow 
(usually called thermocapillary convection) driven by 
the gradient of the surface tension.
The surface tension gradient is indirectly related to the
liquid velocity $v_x$ via the boundary condition 
for the tangential stress at the free surface \cite{levi}:
\begin{equation}
\frac{\partial\gamma}{\partial T}\nabla T=
\eta \frac{\partial v_x}{\partial z} 
\end{equation}
The direction of flow on the liquid surface is expected
to be from hot to cold if the temperature derivative of the tension is
negative. In the opposite case, the surface flow should be
directed toward the hot end. 
Gallium was solidified and transferred to another vacuum      
enclosure having better conditions for creating
a temperature gradient along the water cooled flange.
This chamber was not an UHV: a vacuum of only about 5$\times$10$^{-6}$ Torr could 
be achieved.

 After sputtering the liquid surface we observed the surface flow always to be
 directed from cold to hot end in the middle of the surface.
The flow was visible  by means
of light scattered from the beam footprint on the  surface. Due to the extremely
small scattering cross-section such observations are only possible when
oxide particles occasionally flow through the beam footprint.  
This explaines that $\gamma$ {\it increases} with T in the temperature
range 15 to 35 $^\circ$C. Further heating caused deterioration
of the vacuum with the subsequent oxidation of the sample and quick suppression
of convective flow. Thermocapillary convection usually couples to buoyancy
convection; therefore this reverse direction of the surface flow was
observed only on shallow liquid samples. For deeper layers buoyancy
convection takes over and surface flow is directed from hot to cold regardless
of the temperature behavior of the tension.

In sum, we studied the free Ga surface
using light scattering from capillary waves. Peak frequency of waves
exhibits the temperature-dependent behavior $\omega(T)$
that corresponds to an increase in surface tension up to 100 $^\circ$C
and a decrease in $\gamma$ upon heating to 160 $^\circ$C.
This phenomenon clearly indicates that the surface excess
entropy switches from negative to positive at temperature near 100 
$^\circ$C. The present study shows a clear difference in 
the temperature behavior of tension between liquid gallium and mercury.   
$\partial\gamma/\partial T$ of mercury, studied
with light scattering technique, is continuously
negative in the T-range from 0 to 100 $^\circ$C that is probably due to
a higher effect of capillary waves on the surface excess entropy in
mercury compared to the same contribution in liquid gallium. 

\section*{Acknowledgments} The author is particularly grateful to Professor
W. Freyland of the University of Karlsruhe for stimulating discussions
and critical reading of the manuscript.

\pagebreak
\section*{Captions}
\begin{itemize}
\item[Fig. 1]
Temperature variations of peak frequencies  
of capillary waves of different wavenumbers 
given at each plot in cm$^{-1}$
at the free surface of gallium. The errors in $\omega_0$ are
less than the size of the data points. The solid lines are the
best-fit solutions in the form of Eq. (\ref{slope}).
The experimental points in Fig 1 (b)
were collected on heating (+) and cooling ($\times$). 
\end{itemize}
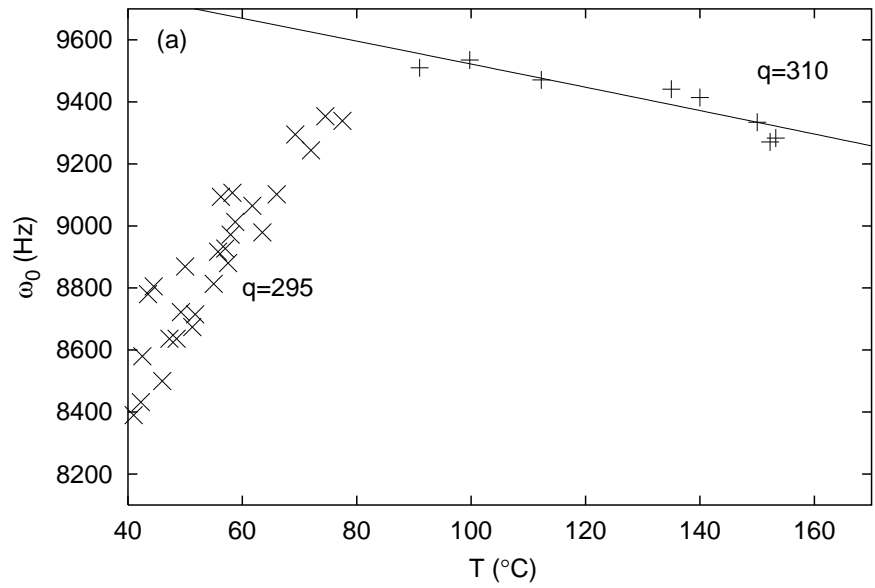
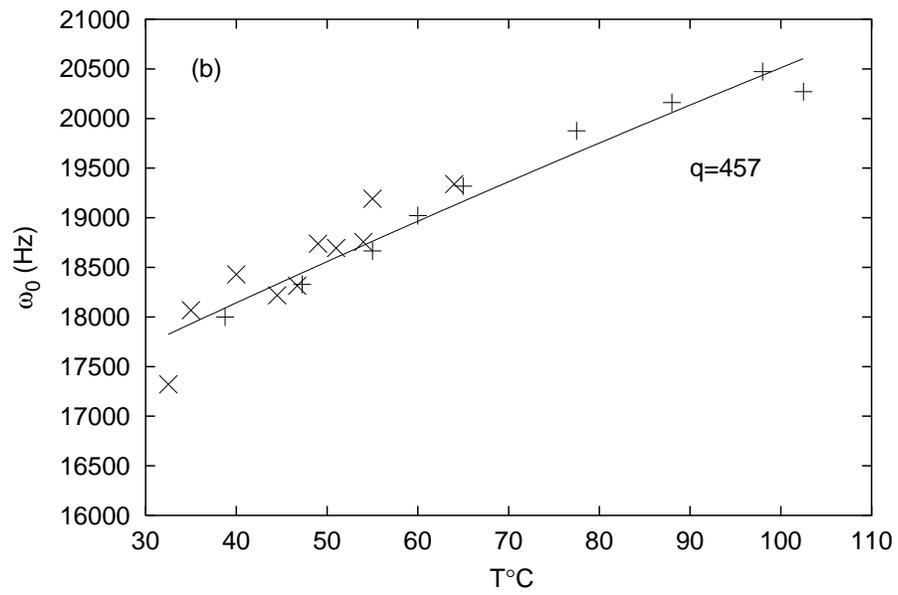
\end{document}